\documentclass[12pt]{article}

\begin{document}

\title{\textbf{Tilted Cylindrically Symmetric Self-Similar Solutions}}

\author{M. Sharif \thanks{msharif@math.pu.edu.pk} and Sajid
Sultan\thanks{sajidsultan@gmail.com}\\
\\Department of Mathematics, University of the Punjab,\\Quaid-e-Azam
Campus Lahore-54590, Pakistan.}
\date{}

\maketitle
\begin{abstract}
This paper is devoted to explore tilted kinematic self-similar
solutions of the the general cylindrical symmetric spacetimes. These
solutions are of the first, zeroth, second and infinite kinds for
the perfect fluid and dust cases. Three different equations of state
are used to obtain these solutions. We obtain a total of five
independent solutions. The correspondence of these solutions with
those already available in the literature is also given.
\end{abstract}


\section{Introduction}

Einstein's theory of General Relativity (GR) relates the geometry
(curvature) to the physical content of spacetime (matter) through
the Einstein field equations (EFEs) given by
\begin{equation}\label{a}
R_{ab}-\frac{1}{2}Rg_{ab}=\kappa T_{ab},
\end{equation}
where $R_{ab}$ is the Ricci tensor, $R$ is the Ricci scalar,
$\kappa$ is the coupling constant and $T_{ab}$ is the
energy-momentum tensor. These equations are coupled, second order,
non-linear, partial differential equations (PDEs) and have no
general solution.

Self-similarity is a scale transformation under which the set of
field equations remains invariant with the assumption of
appropriate matter field. This leads to the existence of
scale-invariant solutions of the field equations which are called
self-similar solutions. The beauty of this restriction is that it
reduces the EFEs to a set of ordinary differential equations
(ODEs) which makes them easier to solve.

Cahill and Taub \cite{C1} were the pioneers who introduced the
concept of self-similarity in GR corresponding to Newtonian
self-similarity of the homothetic class. They studied spherically
symmetric self-similar solutions of the EFEs for a perfect fluid in
the cosmological context. Carr and Coley \cite{C2} discussed
different types of self-similarity in GR. They mainly focussed on
spatially homogenous and spherically symmetric self-similar
solutions. Carr et al. \cite{C3} considered kinematics self-similar
(KSS) vector associated with the critical behavior observed in the
gravitational collapse of spherically symmetric perfect fluid with
equation of state $p = k\rho$. This work was extended \cite{C4} to
discuss the physical aspects of the solutions.

Sintes et al. \cite{C5} investigated KSS solutions of the infinite
kind in the cases of plane, spherically or hyperbolically symmetric
spacetime. Bicknell and Henriksen \cite{C6} considered the
self-similar growth of black holes for a class of equation of state
$p=c^2_s \rho$, where $c_s$ is the constant sound speed. They
concluded that this growth is possible in the sufficiently bounded
regions surrounding the black hole. Mitsuda and Tomimatsu \cite{C7}
investigated the stability of self-similar solutions of
gravitational collapse from the perspective of their nature as an
attractor. They studied the critical phenomena and stability of a
naked singularity. Harada and Maeda \cite{C8} explored spherical
collapse of a perfect fluid with equation of state $p=k\rho$ by full
general relativistic numerical simulations. Ori and Piran \cite{C9}
described a family of general relativistic solutions for
self-similar spherical collapse of an adiabatic perfect fluid
including naked singularities. A counterexample to the
cosmic-censorship hypothesis was also provided. In another paper,
they \cite{C10} examined the structure of general relativistic
spherical collapse solution for a perfect fluid with a barotropic
equation of state.

Sharif and Aziz \cite{C11}-\cite{C12} discussed perfect fluid and
dust solutions for the special and the most general plane symmetric
spacetimes. They explored the first, second, zeroth and infinite
kinds with different equations of state when the KSS vector is
tilted, orthogonal and parallel to the fluid flow. The same authors
\cite{C13}-\cite{C15} discussed the properties of the self-similar
solutions of the first kind for spherically, cylindrically and plane
symmetric spacetimes. They have also studied the self-similar
solutions for a special cylindrically symmetric spacetime
\cite{C16}. Recently, self-similar solutions of the general
cylindrically symmetric spacetime have been investigated \cite{C17}
for the parallel and orthogonal cases. In this paper, we take the
most general cylindrically symmetric spacetimes and find the KSS
solutions of the first, zeroth, second and infinite kinds when the
KSS vector is tilted to the fluid flow both for perfect fluid and
dust cases.

The paper is organized as follows. Section \textbf{2} is devoted in
formulation of the KSS vector for different kinds of
self-similarity. In section \textbf{3} and \textbf{4}, we
investigate self-similar solutions for the tilted perfect fluid and
dust cases respectively. Finally, Section \textbf{5} is furnished
with summary of the results obtained.

\section{Cylindrically Symmetric Spacetimes and Kinematic Self-Similarity}

The line element for the general cylindrically symmetric spacetimes
is given as \cite{C18}
\begin{equation}\label{d11}
ds^{2}=e^{2\nu(t,r)}dt^{2}-e^{2\phi(t,r)}dr^{2}-e^{2\mu(t,r)}d\theta^{2}
-e^{2\lambda(t,r)}dz^{2},
\end{equation}
where $\nu,~ \phi,~ \mu$ and $\lambda$ are functions of $t$ and $r$
only. Notice that the four structure functions are used since we are
using a co-moving frame. The energy-momentum tensor for a perfect
fluid is given by
\begin{equation}
T_{ab}=[\rho(t,r)+p(t,r)]u_{a}u_{b}- p(t,r)g_{ab},\quad(a,b=0, 1,
2, 3),
\end{equation}
where $\rho$ and $p$ are the density and pressure respectively and
$u_{a}$ is the $4$-velocity of the fluid. In co-moving coordinate
system, the $4$-velocity can be written as $u_{a}=(e^{\nu}, 0, 0,
0)$. For the line element (\ref{d11}), the EFEs take the form
\begin{eqnarray}
8\pi
G\rho&=&e^{-2\nu}(\mu_{t}\phi_{t}+\lambda_{t}\phi_{t}+\lambda_{t}\mu_{t})+
e^{-2\phi}(-\mu_{rr}+\mu_{r}\phi_{r}\nonumber\\
&-&\mu^{2}_{r}-\lambda_{rr}+\lambda_{r}\phi_{r}-
\lambda^{2}_{r}-\mu_{r}\lambda_{r}),\\
0&=&-\mu_{tr}-\lambda_{tr}+\mu_{t}\nu_{r}+\lambda_{t}\nu_{r}+\phi_{t}\mu_{r}+
\lambda_{r}\phi_{t}\nonumber\\
&-&\lambda_{r}\lambda_{t}-\mu_{t}\mu_{r},\\
8\pi
Gp&=&e^{-2\nu}(-\mu_{tt}-\lambda_{tt}+\mu_{t}\nu_{t}+\lambda_{t}\nu_{t}-
\lambda_{t}\mu_{t}-\mu^{2}_{t}-\lambda^{2}_{t})\nonumber\\
&+&e^{-2\phi}(\mu_{r}\nu_{r}+\lambda_{r}\nu_{r}+\lambda_{r}\mu_{r}),\\
8\pi
Gp&=&e^{-2\nu}(-\phi_{tt}-\lambda_{tt}+\nu_{t}\phi_{t}+\lambda_{t}\nu_{t}-
\lambda_{t}\phi_{t}-\phi^{2}_{t}-\lambda^{2}_{t})\nonumber\\
&+&e^{-2\phi}(\nu_{rr}+\nu^{2}_{r}-\nu_{r}\phi_{r}+\lambda_{r}\nu_{r}
+\lambda_{rr}-\lambda_{r}\phi_{r}+\lambda^{2}_{r}),\\
8\pi
Gp&=&e^{-2\nu}(-\phi_{tt}-\mu_{tt}+\nu_{t}\phi_{t}+\mu_{t}\nu_{t}-
\mu_{t}\phi_{t}-\phi^{2}_{t}-\mu^{2}_{t})\nonumber\\
&+&e^{-2\phi}(\nu_{rr}+\nu^{2}_{r}-\nu_{r}\phi_{r}+\mu_{r}\nu_{r}
+\mu_{rr}-\mu_{r}\phi_{r}+\mu^{2}_{r}).
\end{eqnarray}
The conservation of energy-momentum tensor, ${T^{ab}}_{;b}=0,$
yields the following equations
\begin{eqnarray}
\phi_{t}=-\frac{\rho_{t}}{\rho+p}-\mu_{t}-\lambda_{t},\quad
\nu_{r}=-\frac{p_{r}}{\rho+p}.
\end{eqnarray}
For a cylindrically symmetric spacetime, the vector field $\xi$ can
have the following form
\begin{equation}
\xi^{a}\frac{\partial}{\partial
x^{a}}=h_{1}(t,r)\frac{\partial}{\partial
t}+h_{2}(t,r)\frac{\partial}{\partial r},
\end{equation}
where $h_{1}$ and $h_{2}$ are arbitrary functions of $t$ and $r$.
The tilted perfect fluid has both $h_1$ and $h_{2}$ non-zero while
$h_1=0$ gives orthogonal case and $h_{2}=0$ the parallel case. This
paper is devoted to investigate the KSS solutions for the tilted
perfect fluid and dust cases.

A kinematic self-similar vector $\xi$ is defined by
\begin{eqnarray}\label{a1}
\pounds_\xi h_{ab}=2 \delta h_{ab},\quad \pounds_\xi u_a=\alpha u_a,
\end{eqnarray}
where $h_{ab}=g_{ab}-u_a u_b$ is the projection tensor and
$\alpha,~\delta$ are dimensionless constants. We can have different
kinds of self-similarity according as $\delta\neq0$ or $\delta=0$
given below:\\
\textbf{$(*)~\delta\neq0$:} Here the KSS vector for the tilted
perfect fluid case takes the form
\begin{equation}\label{st21}
\xi^a\frac{\partial}{\partial x^a}=(\alpha
t+\beta)\frac{\partial}{\partial t}+r\frac{\partial}{\partial r}.
\end{equation}
The similarity index, $\frac{\alpha}{\delta}$, yields the
following three possibilities
\begin{description}
\item[(i)] $\alpha=1$ ($\beta$ can be taken to zero) - first kind,
\item[(ii)] $\alpha=0$ ($\beta$ can be taken to unity) - zeroth
kind,
\item[(iii)] $\alpha\neq0,1$ ($\beta$ can be taken to zero) - second
kind,
\end{description}
where $\delta$ can be taken as unity. The self-similar variable for
self-similarity of the first kind turns out to be $\xi=\frac{r}{t}$.
In the zeroth kind, i.e., $\alpha=0$, the self-similar variable is
$\xi=\frac{r}{e^t}$. For the second kind, the self similar variable
becomes $\xi=r/(\alpha t)^\frac{1}{\alpha}$. For all these kinds,
the metric functions are
\begin{equation}
\nu(t,r)=\nu(\xi),\quad\phi(t,r)=\phi(\xi),\quad
e^{\mu(t,r)}=re^{\mu(\xi)},\quad
e^{\lambda(t,r)}=re^{\lambda(\xi)}.
\end{equation}
\textbf{$(**)~\delta=0$:} Here the KSS vector can take the following
form (when $\alpha\neq0$)
\begin{equation}\label{st22}
\xi^a\frac{\partial}{\partial x^a}=t \frac{\partial}{\partial t}
+r\frac{\partial}{\partial r}
\end{equation}
and the corresponding self-similar variable is $\xi=\frac{r}{t}$.
The metric function will become
\begin{equation}
\nu(t,r)=\nu(\xi),\quad\phi(t,r)=-\ln r+
\phi(\xi),\quad\mu(t,r)=\mu(\xi),\quad\lambda(t,r)=\lambda(\xi).
\end{equation}
The following equations of state (EOS) are used.\\
EOS(1): $p=k\rho^{\gamma}$, where $k$ and $\gamma$ are constants.\\
EOS(2): $p=kn^{\gamma},~\rho=m_{b}n+\frac{p}{\gamma-1}$, where
$k\neq0$ and $\gamma\neq0,1$.\\
EOS(3): $p=k\rho,~-1\leq k \leq 1,~k\neq0$.

\section{Tilted Perfect Fluid Case}

\subsection{Self-similarity of the First Kind}

Using Eq.(13) in Eqs.(4) and (6)-(8), the mass density and pressure
must take the following forms \cite{C19}
\begin{eqnarray}
\kappa \rho(t,r)=\frac{1}{r^2}\rho(\xi),\\
\kappa p(t,r)=\frac{1}{r^2}p(\xi),
\end{eqnarray}
where the self-similar variable is $\xi=r/t$. If the EFEs and the
equations of motion for the matter field are satisfied for
$O[(r)^{-2}]$, we obtain a set of ODEs given by
\begin{eqnarray}
\dot{\rho}&=&-(\dot{\phi}+\dot{\mu}+\dot{\lambda})(\rho+p),\\
2p-\dot{p}&=&\dot{\nu}(\rho+p),\\
0&=&\dot{\mu}\dot{\phi}+\dot{\lambda}\dot{\phi}+\dot{\lambda}\dot{\mu},\\
\rho
e^{2\phi}&=&-\ddot{\mu}-\ddot{\lambda}-\dot{\mu}^2-\ddot{\lambda}^2
-2\dot{\mu}-2\dot{\lambda}+2\dot{\phi}\nonumber\\
&+&\dot{\mu}\dot{\phi}+
\dot{\lambda}\dot{\phi}-\dot{\lambda}\dot{\mu}-1,\\
0&=&\ddot{\mu}+\ddot{\lambda}+\dot{\mu}^2+\ddot{\lambda}^2
+\dot{\mu}+\dot{\lambda}-\dot{\mu}\dot{\nu}-\dot{\lambda}\dot{\nu}\nonumber\\
&-&2\dot{\phi}
-\dot{\phi}\dot{\mu}-\dot{\phi}\dot{\lambda},\\
0&=&-\ddot{\mu}-\ddot{\lambda}-\dot{\mu}^2-\ddot{\lambda}^2
-\dot{\mu}-\dot{\lambda}-\dot{\lambda}\dot{\mu}+\dot{\mu}\dot{\nu}+\dot{\lambda}\dot{\nu},\\
pe^{2\phi}&=&1+\dot{\mu}+\dot{\lambda}+2\dot{\nu}+
\dot{\mu}\dot{\nu}+\dot{\lambda}\dot{\nu}+\dot{\lambda}\dot{\mu},\\
0&=&-\ddot{\phi}-\ddot{\lambda}-\dot{\phi}^2-\ddot{\lambda}^2
-\dot{\phi}-\dot{\lambda}-\dot{\phi}\dot{\lambda}+
\dot{\phi}\dot{\nu}+\dot{\lambda}\dot{\nu},\\
pe^{2\phi}&=&\ddot{\nu}+\ddot{\lambda}+\dot{\nu}^2
+\dot{\lambda}^2+\dot{\lambda}+\dot{\lambda}\dot{\nu}-
\dot{\phi}\dot{\nu}-\dot{\phi}\dot{\lambda}-\dot{\phi},\\
0&=&-\ddot{\phi}-\ddot{\mu}-\dot{\phi}^2-\ddot{\mu}^2
-\dot{\phi}-\dot{\mu}-\dot{\phi}\dot{\mu}+
\dot{\phi}\dot{\nu}+\dot{\mu}\dot{\nu},\\
pe^{2\phi}&=&\ddot{\nu}+\ddot{\mu}+\dot{\nu}^2+\dot{\mu}^2+\dot{\mu}+\dot{\mu}\dot{\nu}-
\dot{\phi}\dot{\nu}-\dot{\phi}\dot{\mu}-\dot{\phi}.
\end{eqnarray}
where dot represents derivative with respect to $\xi$. Sine the
first kind is not compatible with EOS(1) and EOS(2), hence no
solution exists.

\subsubsection{EOS(3)}

If a perfect fluid satisfies EOS(3), then we have the following
solution
\begin{eqnarray}
\nu&=&\ln(c_0\xi^{1\pm\sqrt{2}}),\quad\phi
=c_1,\quad\mu=c_2,\quad\lambda=c_3,\nonumber\\
p&=&\rho=constant
\end{eqnarray}
corresponding to the metric
\begin{equation}\label{aa21}
ds^2=(r)^{2\pm2\sqrt{2}}dt^2-dr^2-r^2(d\theta^2+dz^2).
\end{equation}

\subsection{Self-similarity of the Zeroth Kind}

Here the quantities $\rho$ and $p$ take the form
\begin{eqnarray}
\kappa \rho&=&\frac{1}{r^2}\{\rho_1(\xi)+r^2\rho_2(\xi)\},\\
\kappa p&=&\frac{1}{r^2}\{p_1(\xi)+r^2p_2(\xi)\},
\end{eqnarray}
where the self-similar variable is $\xi=re^{-t}$ and the set of ODEs
become
\begin{eqnarray}
\dot{\rho_1}&=&-(\dot{\phi}+\dot{\mu}+\dot{\lambda})(\rho_1+p_1),\\
\dot{\rho_2}&=&-(\dot{\phi}+\dot{\mu}+\dot{\lambda})(\rho_2+p_2),\\
2p_1-\dot{p_1}&=&\dot{\nu}(\rho_1+p_1),\\
-\dot{p_2}&=&\dot{\nu}(\rho_2+p_2),\\
\rho_1e^{2\phi}&=&-\ddot{\mu}-\ddot{\lambda}-\dot{\mu}^2-\ddot{\lambda}^2
-2\dot{\mu}-2\dot{\lambda}+2\dot{\phi}\nonumber\\
&+&\dot{\mu}\dot{\phi}+
\dot{\lambda}\dot{\phi}-\dot{\lambda}\dot{\mu}-1,\\
\rho_2e^{2\nu}&=&\dot{\mu}\dot{\phi}+\dot{\lambda}\dot{\phi}+\dot{\lambda}\dot{\mu},\\
0&=&\ddot{\mu}+\ddot{\lambda}+\dot{\mu}^2+\ddot{\lambda}^2
+\dot{\mu}+\dot{\lambda}-\dot{\mu}\dot{\nu}-\dot{\lambda}\dot{\nu}\nonumber\\
&-&2\dot{\phi} -\dot{\phi}\dot{\mu}-\dot{\phi}\dot{\lambda},\\
p_1e^{2\phi}&=&1+\dot{\mu}+\dot{\lambda}+2\dot{\nu}+
\dot{\mu}\dot{\nu}+\dot{\lambda}\dot{\nu}+\dot{\lambda}\dot{\mu},\\
p_2e^{2\nu}&=&-\ddot{\mu}-\ddot{\lambda}-\dot{\mu}^2-\ddot{\lambda}^2
-\dot{\lambda}\dot{\mu}+\dot{\mu}\dot{\nu}+\dot{\lambda}\dot{\nu},\\
p_1e^{2\phi}&=&\ddot{\nu}+\ddot{\lambda}+\dot{\nu}^2+\dot{\lambda}^2
+\dot{\lambda}+\dot{\lambda}\dot{\nu}
-\dot{\phi}\dot{\nu}-\dot{\phi}\dot{\lambda}-\dot{\phi},\\
p_2e^{2\nu}&=&-\ddot{\phi}-\ddot{\lambda}-\dot{\phi}^2-\ddot{\lambda}^2
-\dot{\phi}\dot{\lambda}+
\dot{\phi}\dot{\nu}+\dot{\lambda}\dot{\nu},\\
p_1e^{2\phi}&=&\ddot{\nu}+\ddot{\mu}+\dot{\nu}^2+\dot{\mu}^2+\dot{\mu}+\dot{\mu}\dot{\nu}-
\dot{\phi}\dot{\nu}-\dot{\phi}\dot{\mu}-\dot{\phi},\\
p_2e^{2\nu}&=&-\ddot{\phi}-\ddot{\mu}-\dot{\phi}^2-\ddot{\mu}^2
-\dot{\phi}\dot{\mu}+ \dot{\phi}\dot{\nu}+\dot{\mu}\dot{\nu}.
\end{eqnarray}

\subsubsection{Equations of State}

For EOS(1) with $k\neq0$ and $\gamma\neq0,1$, Eqs.(31) and (32)
become
\begin{equation}\label{aa34}
\rho_1=p_1=0,\quad
p_2=\frac{k}{\kappa^{\gamma-1}}\rho^\gamma_2.\quad [Case I]
\end{equation}
For EOS(2) with $k\neq0$ and $\gamma\neq0,1$, Eqs.(31) and (32) take
the form
\begin{equation}\label{aa36}
\rho_1=0=p_1,\quad
p_2=\frac{k}{\kappa^{\gamma-1}m^\gamma_b}(\rho_2-\frac{1}{\gamma-1}p_2)^\gamma.\quad
[Case II]
\end{equation}
EOS(3) yields two more cases for $k=-1$ [Case $III$] and for
$k\neq-1$ [Case $IV$].\\
\textbf{Case I:} Solving the equations simultaneously, we obtain
\begin{eqnarray}
\nu&=&c_1,~\phi=-\ln\xi+\ln(\xi^3-c_3)+c_2,~\mu=-\ln\xi+c_4,~
\lambda=-\ln\xi+c_5,\nonumber\\
p_1&=&0=\rho_1,\quad
p_2=constant,\quad\rho_2=\frac{-3(\xi^3+c_2)}{e^{2c_1}(\xi^3-c_3)}.
\end{eqnarray}
The corresponding metric is
\begin{equation}\label{aa39}
ds^2=dt^2-(\frac{r^3-c_3e^{3t}}{re^{2t}})^2dr^2-e^{2t}(d\theta^2+dz^2).
\end{equation}
The second solution is
\begin{eqnarray}
\nu&=&c_1,\quad\phi=\frac{1}{2}\ln\xi+c_2,\quad
\mu=\frac{1}{2}\ln\xi+c_3,\quad\lambda=-\ln\xi+c_4,\nonumber\\
\rho_1&=&0=p_1,\quad \rho_2=p_2=constant
\end{eqnarray}
and the corresponding metric is
\begin{equation}\label{aa45}
ds^2=dt^2-\frac{r}{e^t}(dr^2+r^2d\theta^2)-e^{2t}dz^2.
\end{equation}
The third solution is
\begin{eqnarray}
\nu&=&c_1,\quad\phi=\frac{1}{2}\ln\xi+c_2,\quad
\mu=-\ln\xi+c_3,\quad\lambda=\frac{1}{2}\ln\xi+c_4,\nonumber\\
\rho_1&=&0=p_1,\quad \rho_2=p_2=constant
\end{eqnarray}
and the metric takes the form
\begin{equation}\label{aa46}
ds^2=dt^2-\frac{r}{e^t}dr^2-e^{2t}d\theta^2-\frac{r^3}{e^t}dz^2.
\end{equation}
The Case \textbf{II} gives the same solutions as the Case \textbf{I}.\\
\textbf{Case III:} Here the solution becomes
\begin{eqnarray}
\nu&=&c_1,\quad\phi=-\ln\xi+c_2,\quad\mu=-\ln\xi+c_3,\quad
\lambda=-\ln\xi+c_4,\nonumber\\
\rho_1&=&0=p_1,\quad\rho_2=p_2=constant
\end{eqnarray}
and the corresponding metric is
\begin{equation}\label{aa47}
ds^2=dt^2-\frac{e^{2t}}{r^2}(dr^2+r^2(d\theta^2+dz^2)).
\end{equation}
\textbf{Case IV}: This case gives the two solutions out of which the
first is
\begin{eqnarray}
\nu&=&c_1,\quad\phi=2\ln\xi+c_2,\quad
\mu=-\ln\xi+c_3,\quad\lambda=-\ln\xi+c_4,\nonumber\\
\rho_1&=&0=p_1,\quad\rho_2=p_2=constant
\end{eqnarray}
and the corresponding metric is
\begin{equation}\label{aa54}
ds^2=dt^2-\frac{r^4}{e^{4t}}dr^2-e^{2t}(d\theta^2+dz^2).
\end{equation}
The second solution is
\begin{eqnarray}
\nu&=&(1\pm\sqrt{2})\ln\xi+c_1,\quad\phi=c_2,\quad\mu=c_3,\quad
\lambda=c_4,\nonumber\\
\rho_1&=&p_1=constant,\quad\rho_2=0=p_2
\end{eqnarray}
and the corresponding spacetime is
\begin{equation}\label{aa56}
ds^2=(\frac{r}{e^t})^{2\pm2\sqrt{2}}dt^2-dr^2-r^2(d\theta^2+dz^2).
\end{equation}

\subsection{Self-similarity of the Second Kind}

Here the EFEs imply that that
\begin{eqnarray}
\kappa \rho&=&\frac{1}{r^2}\{\rho_1(\xi)+\frac{r^2}{t^2}\rho_2(\xi)\},\\
\kappa p&=&\frac{1}{r^2}\{p_1(\xi)+\frac{r^2}{t^2}p_2(\xi)\},
\end{eqnarray}
where the self-similar variable is $\xi=r/(\alpha t)^{1/\alpha}$.
The set of ODEs become
\begin{eqnarray}
\dot{\rho_1}&=&-(\dot{\phi}+\dot{\mu}+\dot{\lambda})(\rho_1+p_1),\\
\dot{\rho_2}+2\alpha
\rho_2&=&-(\dot{\phi}+\dot{\mu}+\dot{\lambda})(\rho_2+p_2),\\
2p_1-\dot{p_1}&=&\dot{\nu}(\rho_1+p_1),\\
-\dot{p_2}&=&\dot{\nu}(\rho_2+p_2),\\
\rho_1e^{2\phi}&=&-\ddot{\mu}-\ddot{\lambda}-\dot{\mu}^2-\ddot{\lambda}^2
-2\dot{\mu}-2\dot{\lambda}+2\dot{\phi}\nonumber\\
&+&\dot{\mu}\dot{\phi}+
\dot{\lambda}\dot{\phi}-\dot{\lambda}\dot{\mu}-1,\\
\alpha^2\rho_2e^{2\nu}&=&\dot{\mu}\dot{\phi}+\dot{\lambda}\dot{\phi}+\dot{\lambda}\dot{\mu},\\
0&=&\ddot{\mu}+\ddot{\lambda}+\dot{\mu}^2+\ddot{\lambda}^2
+\dot{\mu}+\dot{\lambda}-\dot{\mu}\dot{\nu}-\dot{\lambda}\dot{\nu}\nonumber\\
&-&2\dot{\phi}
-\dot{\phi}\dot{\mu}-\dot{\phi}\dot{\lambda},\\
p_1e^{2\phi}&=&1+\dot{\mu}+\dot{\lambda}+2\dot{\nu}+
\dot{\mu}\dot{\nu}+\dot{\lambda}\dot{\nu}+\dot{\lambda}\dot{\mu},\\
\alpha^2p_2e^{2\nu}&=&-\ddot{\mu}-\ddot{\lambda}-\dot{\mu}^2-\ddot{\lambda}^2
-\alpha \dot{\mu}-\alpha\dot{\lambda}\nonumber\\
&-&\dot{\lambda}\dot{\mu}+\dot{\mu}\dot{\nu}+\dot{\lambda}\dot{\nu},\\
p_1e^{2\phi}&=&\ddot{\nu}+\ddot{\lambda}+\dot{\nu}^2
+\dot{\lambda}^2+\dot{\lambda}+\dot{\lambda}\dot{\nu}-
\dot{\phi}\dot{\nu}-\dot{\phi}\dot{\lambda}-\dot{\phi},\\
\alpha^2p_2e^{2\nu}&=&-\ddot{\phi}-\ddot{\lambda}-\dot{\phi}^2-\ddot{\lambda}^2-\alpha
\dot{\phi}-\alpha\dot{\lambda}\nonumber\\
&-&\dot{\phi}\dot{\lambda}+
\dot{\phi}\dot{\nu}+\dot{\lambda}\dot{\nu},\\
p_1e^{2\phi}&=&\ddot{\nu}+\ddot{\mu}+\dot{\nu}^2+\dot{\mu}^2+\dot{\mu}+\dot{\mu}\dot{\nu}-
\dot{\phi}\dot{\nu}-\dot{\phi}\dot{\mu}-\dot{\phi},\\
\alpha^2p_2e^{2\nu}&=&-\ddot{\phi}-\ddot{\mu}-\dot{\phi}^2-\ddot{\mu}^2-\alpha
\dot{\phi}-\alpha\dot{\mu}\nonumber\\
&-&\dot{\phi}\dot{\mu}+ \dot{\phi}\dot{\nu}+\dot{\mu}\dot{\nu}.
\end{eqnarray}

\subsubsection{Equations of State}

For EOS(1) ($k\neq0$ and $\gamma\neq0,1$), Eqs.(60) and (61) imply
that
\begin{eqnarray}
p_1&=&0=\rho_2,\quad\alpha=\gamma,\quad
p_2=\frac{k}{\kappa^{\gamma-1}
\gamma^2}\xi^{-2\gamma}\rho^\gamma_1,\quad [Case I]\\
p_2&=&0=\rho_1,\quad\alpha=\frac{1}{\gamma},\quad p_1
=\frac{k}{\kappa^{\gamma-1}\gamma^{2\gamma}}\xi^2\rho^\gamma_2.\quad
[Case II]
\end{eqnarray}
When a perfect fluid satisfies EOS(2) for $k\neq0$ and
$\gamma\neq0,1$, it follows from Eqs.(60) and (61) that
\begin{eqnarray}
p_2&=&\frac{k}{m^\gamma_b\gamma^2\kappa^{\gamma-1}}
\xi^{-2\gamma}\rho^\gamma_1=(\gamma-1)\rho_2,\nonumber\\
p_1&=&0,\quad\alpha=\gamma,\quad [Case III]\\
p_1&=&\frac{k}{m^\gamma_b \gamma^{2\gamma}\kappa^{\gamma-1}}
\xi^2\rho^\gamma_2=(\gamma-1)\rho_1,\nonumber\\
p_2&=&0,\quad\alpha=\frac{1}{\gamma}.\quad[Case IV]
\end{eqnarray}
EOS(3) yields two more cases (Case V and Case VI) for $k=-1$ and
$k\neq-1$ respectively.\\
\textbf{Case I:} Solving the ODEs simultaneously, we obtain the
solutions as
\begin{eqnarray}
\nu&=&c_1,\quad \phi=\frac{1}{2}\ln\xi+c_2,\quad
\mu=-\ln\xi+c_3,\quad\lambda=-\ln\xi+c_4,\nonumber\\
\rho_1&=&\rho_2=0=p_1=p_2,\quad\alpha=\frac{3}{2}
\end{eqnarray}
and the corresponding metric is
\begin{equation}\label{aa67}
ds^2=dt^2-(\frac{2}{3t})^\frac{2}{3}rdr^2
-(\frac{3t}{2})^\frac{4}{3}(d\theta^2+dz^2).
\end{equation}
The second solution is
\begin{eqnarray}
\nu&=&c_1,\quad \phi=2 \ln \xi+c_2,\quad \mu=-\ln \xi+c_3,\quad
\lambda=2 \ln\xi+c_4,\nonumber\\
\rho_1&=&\rho_2=0=p_1=p_2,\quad\alpha=-3.
\end{eqnarray}
Its metric form is
\begin{equation}\label{aa71}
ds^2=dt^2-r^4(-3t)^{\frac{4}{3}}dr^2-\frac{1}{(-3t)^\frac{2}{3}}
d\theta^2-r^6(-3t)^\frac{4}{3}dz^2.
\end{equation}
The third solution is
\begin{eqnarray}
\nu&=&c_1,\quad \phi=2 \ln \xi+c_2,\quad \mu=2 \ln
\xi+c_4,\quad \lambda=-\ln \xi+c_3,\nonumber\\
\rho_1&=&\rho_2=0=p_1=p_2,\quad\alpha=-3
\end{eqnarray}
and the corresponding metric is
\begin{equation}\label{aa72}
ds^2=dt^2-r^4(-3t)^{\frac{4}{3}}(dr^2+r^2
d\theta^2)-\frac{1}{(-3t)^\frac{2}{3}}dz^2.
\end{equation}
It is mentioned here that the Cases II, III and IV have the same
solutions as given by Eqs.({\ref{aa67}), ({\ref{aa71}) and
({\ref{aa72}). It is also noted that the Case V has only
one solution given by Eq.({\ref{aa67}}).\\
\textbf{Case VI:} For this case, we obtain the following solutions.
The first solution is
\begin{eqnarray}
\nu&=&c_1,\quad\phi=-\ln\xi+c_2,\quad
\mu=-\ln\xi+c_3,\quad\lambda=-\ln\xi+c_4,\nonumber\\
\rho_1&=&0=p_1,\quad\rho_2=p_2=constant,\quad
k=\frac{2\alpha-3}{3}
\end{eqnarray}
and the corresponding metric is
\begin{equation}\label{aa85}
ds^2=dt^2-\frac{(\alpha t)^\frac{2}{\alpha}}{r^2}dr^2-(\alpha
t)^\frac{2}{\alpha}(d\theta^2+dz^2).
\end{equation}
The second solution is
\begin{eqnarray}
\nu&=&c_1,\quad\phi=(2-\alpha)\ln\xi=c_2,\quad\mu=-\ln\xi+c_3,\quad
\lambda=-\ln\xi+c_4,\nonumber\\
\rho_1&=&0=p_1,\quad\rho_2=p_2=constant,\quad k=1
\end{eqnarray}
and the corresponding spacetime is
\begin{equation}\label{aa86}
ds^2=dt^2-\frac{r^{2(2-\alpha)}}{(\alpha
t)^\frac{2(2-\alpha)}{\alpha}}dr^2-(\alpha
t)^\frac{2}{\alpha}(d\theta^2+dz^2).
\end{equation}
The third solution is
\begin{eqnarray}
\nu&=&c_1,\quad\phi=-\ln\xi=c_2,\quad
\mu=-\ln\xi+c_3,\quad\lambda=-\ln\xi+c_4,\nonumber\\
\rho_1&=&0=p_1,\quad\rho_2=p_2=constant,\quad k=1,\quad \alpha=3
\end{eqnarray}
and its metric is
\begin{equation}\label{ab90}
ds^2=dt^2-\frac{(3t)^\frac{2}{3}}{r^2}dr^2-(3t)^\frac{2}{3}(d\theta^2+dz^2).
\end{equation}
The fourth solution is
\begin{eqnarray}
\nu&=&(1\pm\sqrt{2})\ln\xi+c_1,\quad
\phi=c_2,\mu=c_3,\quad\lambda=c_4,\nonumber\\
\rho_1&=&p_1=constant,\quad \rho_2=0=p_2,\quad k=-3\pm2\sqrt{2}
\end{eqnarray}
and the corresponding metric is
\begin{equation}\label{ab93}
ds^2=(\frac{r}{(\alpha
t)^{1/\alpha}})^{2\pm2\sqrt{2}}dt^2-dr^2-r^2(d\theta^2+dz^2).
\end{equation}

\subsection{Self-similarity of the Infinite Kind}

Here the quantities $\rho$ and $p$ can be written in terms of $\xi$
as
\begin{eqnarray}
\kappa \rho&=&\rho_1(\xi)+\frac{1}{t^2}\rho_2(\xi),\\
\kappa p&=&p_1(\xi)+\frac{1}{t^2}p_2(\xi),
\end{eqnarray}
where the self-similar variable is $\xi=r/t$ and the set of ODEs
become
\begin{eqnarray}
\dot{\rho_1}&=&-(\dot{\phi}+\dot{\mu}+\dot{\lambda})(\rho_1+p_1),\\
\dot{\rho_2}+2\rho_2&=&-(\dot{\phi}+\dot{\mu}+\dot{\lambda})(\rho_2+p_2),\\
-\dot{p_1}&=&\dot{\nu}(\rho_1+p_1),\\
-\dot{p_2}&=&\dot{\nu}(\rho_2+p_2),\\
\rho_1e^{2\phi}&=&-\ddot{\mu}-\ddot{\lambda}-\dot{\mu}^2-\ddot{\lambda}^2
+\dot{\mu}\dot{\phi}+
\dot{\lambda}\dot{\phi}-\dot{\lambda}\dot{\mu},\\
\rho_2e^{2\nu}&=&\dot{\mu}\dot{\phi}+\dot{\lambda}\dot{\phi}+\dot{\lambda}\dot{\mu},\\
0&=&\ddot{\mu}+\ddot{\lambda}+\dot{\mu}^2+\ddot{\lambda}^2
-\dot{\mu}\dot{\nu}-\dot{\lambda}\dot{\nu}
-\dot{\phi}\dot{\mu}-\dot{\phi}\dot{\lambda},\\
p_1e^{2\phi}&=&\dot{\mu}\dot{\nu}+\dot{\lambda}\dot{\nu}+\dot{\lambda}\dot{\mu},\\
p_2e^{2\nu}&=&-\ddot{\mu}-\ddot{\lambda}-\dot{\mu}^2-\ddot{\lambda}^2
-\dot{\mu}-\dot{\lambda}
-\dot{\lambda}\dot{\mu}+\dot{\mu}\dot{\nu}+\dot{\lambda}\dot{\nu},\\
p_1e^{2\phi}&=&\ddot{\nu}+\ddot{\lambda}+\dot{\nu}^2+\dot{\lambda}^2
+\dot{\lambda}\dot{\nu}-\dot{\phi}\dot{\nu}-\dot{\phi}\dot{\lambda},\\
p_2e^{2\nu}&=&-\ddot{\phi}-\ddot{\lambda}-\dot{\phi}^2
-\ddot{\lambda}^2-\dot{\phi}-\dot{\lambda}-\dot{\phi}\dot{\lambda}
+\dot{\phi}\dot{\nu}+\dot{\lambda}\dot{\nu},\\
p_1e^{2\phi}&=&\ddot{\nu}+\ddot{\mu}+\dot{\nu}^2+\dot{\mu}^2+\dot{\mu}\dot{\nu}
-\dot{\phi}\dot{\nu}-\dot{\phi}\dot{\mu},\\
p_2e^{2\nu}&=&-\ddot{\phi}-\ddot{\mu}-\dot{\phi}^2-\ddot{\mu}^2-\dot{\phi}-\dot{\mu}
-\dot{\phi}\dot{\mu}+ \dot{\phi}\dot{\nu}+\dot{\mu}\dot{\nu}.
\end{eqnarray}

\subsubsection{Equations of State}

For EOS(1) with $k\neq0$ and $\gamma\neq0,1$, Eqs.(95) and (96)
become
\begin{equation}\label{aa94}
\rho_2=0=p_2,\quad
p_1=\frac{k}{\kappa^{\gamma-1}}\rho^\gamma_1.\quad[Case~I]
\end{equation}
For EOS(2) with $k\neq0$ and $\gamma\neq0,1$, Eqs.(95) and (96) take
the form
\begin{equation}\label{aa96}
\rho_2=0=p_2,\quad p_1=\frac{k}{m^\gamma_b\kappa^{\gamma-1}}
(\rho_1-\frac{p_1}{\gamma-1})^\gamma.\quad[Case~ II]
\end{equation}
EOS(3) yields two more cases for $k=-1$ [Case $III$] and for
$k\neq-1$ [Case $IV$].\\
\textbf{Case I}: Solving the ODEs simultaneously for this case, we
have
\begin{eqnarray}
\nu=c_1,\quad\phi=c_2,\quad\mu=c_3,\quad\lambda=c_4,\quad
\rho_1=\rho_2=0=p_1=p_2
\end{eqnarray}
and the corresponding metric is
\begin{equation}\label{ts15}
ds^2=dt^2-\frac{1}{r^2}dr^2-d\theta^2-dz^2
\end{equation}
which corresponds to Minkowski spacetime. The second solution is
\begin{eqnarray}
\nu&=&\ln(\ln\xi-\ln c_1)+c_2,\quad\phi=c_3,\quad\mu=c_4,\quad\lambda=c_5,\nonumber\\
\rho_1&=&\rho_2=0=p_1=p_2.
\end{eqnarray}
The corresponding metric is
\begin{equation}\label{aa99}
ds^2=[\ln(\frac{r}{c_1t})]^2dt^2-\frac{1}{r^2}dr^2-(d\theta^2+dz^2).
\end{equation}
The third solution becomes
\begin{eqnarray}
\nu=c_3,\quad\phi=\ln(\frac{\xi-c_1}{\xi})+c_2,\quad
\dot{\mu}=0,~\dot{\lambda}=0,~ \rho_1=\rho_2=0=p_1=p_2.
\end{eqnarray}
The corresponding metric is
\begin{equation}\label{ss00}
ds^2=dt^2-\frac{(r-c_1t)^2}{r^4}dr^2-(d\theta^2+dz^2).
\end{equation}
It is mentioned here that the Cases II, III and IV have the same
solutions as the Case I.

\section{Tilted Dust Case}

It is well-known that a perfect fluid is characterized by the
pressure and it reduces to the dust fluid when $p=0$.

\subsection{Self-similarity of the First Kind}

When we substitute $p=0$ in Eqs.(18)-(28), it follows from Eq.(19)
\begin{equation}\label{ss11}
\dot{\nu}\rho=0
\end{equation}
which gives the following two solutions. The first solution turns
out to be
\begin{equation} \nu=c_1,\quad
\phi=c_2,\quad \mu=c_3,\quad \lambda=-\ln \xi+c_4,\quad \rho=0
\end{equation}
with metric
\begin{equation}\label{ss17}
ds^2=dt^2-dr^2-r^2d\theta^2-t^2dz^2.
\end{equation}
The second solution is
\begin{equation}
\nu=c_1,\quad \phi=c_2,\quad \mu=-\ln\xi+c_3,\quad
\lambda=c_4,\quad \rho=0
\end{equation}
and the corresponding metric is
\begin{equation}\label{ss18}
ds^2=dt^2-dr^2-t^2d\theta^2-r^2dz^2.
\end{equation}

\subsection{Self-similarity of the Zeroth Kind}

When we take $p_1=0$ and $p_2=0$ in Eqs.(33)-(45), we obtain
contradiction and consequently there is no self-similar solution of
the zeroth kind.

\subsection{Self-similarity of the Second Kind}

When we substitute $p_1=0$ and $p_2=0$ in Eqs.(62)-(74), we obtain
the following solution
\begin{eqnarray}
\nu&=&c_1,\quad \phi=\ln(\frac{2\xi^\frac{3}{2}-c_2}{\xi}),\quad
\mu=-\ln\xi+c_3,\quad \lambda=-\ln\xi+c_4,\nonumber\\
\rho_1&=&0,\quad
\rho_2=\frac{4}{9e^{2c_1}}(\frac{3c_2}{c_2-2\xi^\frac{3}{2}}).
\end{eqnarray}
The corresponding metric is
\begin{equation}\label{ss23}
ds^2=dt^2-(\frac{3t}{2})^\frac{4}{3}(\frac{4r^\frac{3}{2}
-3tc_2}{3rt})^2dr^2-(\frac{3t}{2})^\frac{4}{3}(d\theta^2+dz^2).
\end{equation}

\subsection{Self-similarity of the Infinite Kind}

Taking $p_{1}=0=p_{2}$ in Eqs.(97)-(109), it yields the same
solutions as given by Eqs.({\ref{ts15}}) and ({\ref{ss00}}).

\section{Summary and Discussion}

Recently, Sharif and Aziz \cite{C16} have found the KSS solutions
for a special cylindrically symmetric spacetimes. This work has been
extended to investigate the KSS solutions of the general
cylindrically symmetric spacetime when the fluid flow is parallel as
well as orthogonal \cite{C17}. However, we have left the tilted case
to deal it separately. In this paper, we have studied the case when
KSS vector is tilted to the fluid flow. The solutions are found for
the first, zeroth, second and the infinite kinds both for perfect
fluid and dust cases. The summary of the independent solutions is
given in the form of table \textbf{1}.
\begin{center}
\textbf{Table 1.} Tilted KSS Solutions.\vspace{0.5cm}
\begin{tabular}{|l|l|l|l|}
\hline  {\bf S. NO.} & {\bf Metric} & {\bf Equation of state}& {\bf
Physical}
\\ \hline I& $\begin{array}{c}
ds^2=(r)^{2\pm2\sqrt{2}}dt^2\\-dr^2-r^2(d\theta^2+dz^2)
\end{array}$ & $\begin{array}{c}\rho=-\frac{1}{8\pi r^2}<0,\\
p=-(3\pm2\sqrt{2})\rho>0\end{array}$& Unphysical
\\ \hline  II& $\begin{array}{c}
ds^2=dt^2\\-\frac{r}{e^t}(dr^2+r^2d\theta^2)-e^{2t}dz^2
\end{array}$ &$\begin{array}{c} \rho=p=-\frac{3}{32\pi}<0
\end{array}$&Unphysical
\\ \hline III& $\begin{array}{c}
ds^2=dT^2-e^{2T}(dR^2+R^2d\Theta^2\\+dZ^2),\\
R>0,~-\infty<T<+\infty,\\0<\Theta<2\pi,~-\infty<Z<+\infty
\end{array}$ &$\begin{array}{c} \rho=-p=\frac{3}{8\pi}>0
\end{array}$&Physical
\\ \hline IV& $\begin{array}{c}
ds^2=dt^2-r^4(3t)^{\frac{4}{3}}(dr^2\\+r^2
d\theta^2)-\frac{1}{(3t)^\frac{2}{3}}dz^2
\end{array}$ &$\rho=0=p$ &Physical
\\ \hline V& $\begin{array}{c}
ds^2=dT^2-dR^2-R^2d\Theta^2\\+dZ^2,\\
R>0,~-\infty<T<+\infty,\\0<\Theta<2\pi,~-\infty<Z<+\infty\end{array}$
&$\rho=0=p$ &Physical
\\ \hline
\end{tabular}
\end{center}

The results can be discussed as follows:\\
I. $\rho=-\frac{1}{8\pi r^2}<0,~p=-(3\pm2\sqrt{2})\rho>0$. This
solution is unphysical as energy density is negative.\\
II. For this solution, we have $\rho=p=-\frac{3}{32\pi}<0$ which is
unphysical. \\
III. This is the expanding de Sitter universe (empty expanding flat
space with cosmological constant $\Lambda>0$). In this case $p<0$
can be re-interpreted and considered by most astrophysicists as
physical. \\
IV. Here $\rho=0=p$. This is an empty but curved spacetime as
$R=0,~R_{ab}=0$, but $R_{abcd}R^{abcd}=\frac{64}{27t^4}\neq0$
singular at $t=0$ (extendible). \\
V. This gives $\rho=0=p,~R_{abcd}R^{abcd}=0$, i.e. empty flat,
Minkowski spacetime in cylindrical coordinates.

We have seen that the number of independent solutions are very few.
Also, most of these solutions coincide with those already available
in the literature. Thus we can conclude that one may obtain already
known spacetimes by specializing a map adapted to some symmetries.
However, the line element may look more complicated in such a map.
Moreover, such a map may cover only a fragment of the spacetime.
\vspace{0.5cm}

{\bf Acknowledgment}

\vspace{0.5cm}

We would like to appreciate the referee's useful comments.


\begin{thebibliography}{99}

\bibitem{C1} Cahill, M.E. and Taub, A.H.: Commun. Math. Phys.
\textbf{21}(1971)1.

\bibitem{C2} Carr, B.J. and Coley, A.A.: Class. Quantum Grav.
\textbf{16}(1999)R31.

\bibitem{C3} Carr, B.J., Coley, A.A., Goliath, M., Nilsson, U.S.
and Uggla, C.: Phys. Rev. \textbf{D61}(2000)081502.

\bibitem{C4} Carr, B.J., Coley, A.A., Goliath, M., Nilsson, U.S.
and Uggla, C.: Class. Quantum Grav. \textbf{18}(2001)303.

\bibitem{C5} Sintes, A.M., Benoit, P.M. and Coley, A.A.: Gen. Relat. Gravit.
\textbf{33}(2001)1863.

\bibitem{C6} Bicknell, G.V. and Henriksen, R.N.: Astrophysical J.
\textbf{225}(1978)251.

\bibitem{C7} Mitsuda, E. and Tomimatsu, A.: Phys. Rev.
\textbf{D73}(2006)124030.

\bibitem{C8} Harada, T. and Maeda, H.: Phys. Rev.
\textbf{D63}(2001)084022.

\bibitem{C9} Ori, A. and Piran, T.: Phys. Rev. Lett.
\textbf{59}(1987)2137.

\bibitem{C10} Ori, A. and Piran, T.: Phys. Rev.
\textbf{D42}(1990)1068.

\bibitem{C11} Sharif, M. and Aziz, S.: J. Korean Physical Society
\textbf{49}(2006)21.

\bibitem{C12} Sharif, M. and Aziz, S.: Class. Quantum Grav.
\textbf{24}(2007)605.

\bibitem{C13} Sharif, M. and Aziz, S.: Int. J. Mod. Phys.
\textbf{D14}(2005)73.

\bibitem{C14} Sharif, M. and Aziz, S.: Int. J. Mod. Phys.
\textbf{A20}(2005)7579.

\bibitem{C15} Sharif, M. and Aziz, S.: J. Korean Physical Society
\textbf{47}(2005)757.

\bibitem{C16} Sharif, M. and Aziz, S.: Int. J. Mod. Phys.
\textbf{D14}(2005)1527.

\bibitem{C17} Sharif, M. and Zanub, J.: J. Korean Physical Society (2009 to appear);\\
Zanub, J.: {\it M. Phil. Thesis} (University of the Punjab, Lahore,
2007).

\bibitem{C18} Stephani, H., Kramer, D., MacCallum, M., Hoenselaers, C. and Herlt,
E.: {\it Exact Solutions of Einstein's Field Equations} (Cambridge
University Press, 2003).

\bibitem{C19} Sultan, S.: {\it M. Phil. Thesis} (University of the Punjab, Lahore,
2007).
\end{thebibliography}
\end{document}